\documentclass[sn-mathphys-num]{sn-jnl}

\usepackage{graphicx}%
\usepackage{multirow}%
\usepackage{amsmath,amssymb,amsfonts}%
\usepackage{amsthm}%
\usepackage{mathrsfs}%
\usepackage[title]{appendix}%
\usepackage{xcolor}%
\usepackage{textcomp}%
\usepackage{manyfoot}%
\usepackage{booktabs}%
\usepackage{algorithm}%
\usepackage{algorithmicx}%
\usepackage{algpseudocode}%
\usepackage{listings}%

\theoremstyle{thmstyleone}%
\theoremstyle{thmstyletwo}%

\theoremstyle{thmstylethree}%

\begin{document}

\title[Article Title]{CTMBIDS: Convolutional Tsetlin Machine Based Intrusion Detection System for DDoS attacks in an SDN environment}


\author[1]{\fnm{Rasoul} \sur{Jafari Gohari}}\email{rjafari@math.uk.ac.ir}

\author[2*]{\fnm{Laya} \sur{Aliahmadipour}}\email{l.aliahmadipour@uk.ac.ir}

\author[3]{\fnm{Marjan} \sur{Kuchaki Rafsanjani}}\email{kuchaki@uk.ac.ir}

\affil[1,2*,3]{\orgdiv{Department of Computer Science}, \orgname{Shahid Bahonar University of Kerman}, \orgaddress{\street{Paghohesh Square}, \city{Kerman}, \postcode{7616913439}, \state{Kerman}, \country{Iran}}}


\abstract{Software Defined Networks (SDN) face many security challenges today. A great deal of research has been done within the field of Intrusion Detection Systems (IDS) in these networks. Yet, numerous approaches still rely on deep learning algorithms, but these algorithms suffer from complexity in implementation, the need for high processing power, and high memory consumption. In addition to security issues, firstly, the number of datasets that are based on SDN protocols are very small. Secondly, the ones that are available encompass a variety of attacks in the network and don't focus on a single attack. For this reason, to introduce an SDN-based IDS with a focus on Distributed Denial of Service (DDoS) attacks, it is necessary to generate a DDoS-oriented dataset whose features can train a high-quality IDS. In this work, in order to address two important challenges in SDNs, in the first step, we generate three DDoS attack datasets based on three common and different network topologies. Then, in the second step, using the Convolutional Tsetlin Machine (CTM) algorithm, we introduce a lightweight IDS for DDoS attack dubbed "CTMBIDS", with which we implement an anomaly-based IDS. The lightweight nature of the CTMBIDS stems from its low memory consumption and also its interpretability compared to the existing complex deep learning models. The low usage of system resources for the CTMBIDS makes it an ideal choice for an optimal software that consumes the SDN controller's least amount of memory. Also, in order to ascertain the quality of the generated datasets, we compare the empirical results of our work with the DDoS attacks of the KDDCup99 benchmark dataset as well. Since the main focus of this work is on a lightweight IDS, the results of this work show that the CTMBIDS performs much more efficiently than traditional and deep learning based machine learning algorithms. Furthermore, the results also show that in most datasets, the proposed method has relatively equal or better accuracy and also consumes much less memory than the existing methods. }




\maketitle

\section{Introduction}\label{sec1}

The shortcomings of conventional networks have recently made room for the creation of Software Defined Network (SDN). This evolution today is considered a milestone and has diversified itself into other important fields such as SD-Wide Area Networks (SD-WAN) \cite{1}, SDN-Internet of Things (SDNIoT) \cite{2}, SD5g and SD6g \cite{3}. This is due to the fact that SDN architecture allows for the separation of data layer and control layer in the network \cite{4}, which as a result yields a very functional environment, in which network administrators can work very efficiently. This efficiency stems from the flexibility, programmability and manageability of SDN architecture that has roots in the decoupling of data layer and control layer and centralizing the network management in an SDN controller. However, with all the benefits that SDN architecture puts forward, SDN is still susceptible to a wide range of network attacks. Therefore, network attacks and their detection is still considered an urgent need. Numerous methods have been proposed to combat this challenge, one of which is the implementation of Intrusion Detection System (IDS) in the network. The idea of an IDS in the computer networks becomes even more functional when combined with the programmability side of SDN architecture. In this way, not only network administrators can easily implement an IDS in the control layer part of the SDN architecture to have a comprehensive view of the network, but also they lower the costs substantially via removing all the vendor-based IDSs from the network \cite{5}. Moreover, an IDS in conventional networks can only detect attacks as far as its access and observability is concerned while SDN flexibility gives room to an IDS to be far more advanced in terms of access and comprehensive view of the entire network. 

Network Intrusion Detection Systems (NIDS) are considered the ideal approach to perform intrusion detection across the network \cite{6}. The fundamental utility of a NIDS is to monitor the traffic throughout the network and detect any intrusion that occurs so it can report it to network administrators. NIDSs are normally categorized into signature-based and anomaly-based intrusion detection systems. Signature-based NIDS is capable of detecting intrusions based on a body of knowledge stored in a database as signatures of previous attacks. This means the slightest shift in the behavior of the intrusion will allow the attacker to bypass the signature-based IDS. Furthermore, keeping the database of the attack signatures updated is another major hurdle that can become an arduous task when the network administrators have to take the storage of the signatures into account.  Anomaly-based IDSs on the other hand are more suited for unknown attacks \cite{7}. The advantage of such IDS is substantially increasing the scope of intrusion detections such that newly implemented attacks that deviate from the signatures of previously-implemented intrusions are far more easily detectable. Machine Learning (ML) algorithms enable IDS to detect specific patterns and anomalies in the network. The main focus of the research community is primarily on anomaly-based IDS since the complex patterns and new methodologies in different types of attacks can be easily discovered. 

An applicable ML algorithm that normally induces better results in terms of accuracy and low cost is Deep Learning (DL) algorithm. Not only DL-based models are extremely efficient for high-dimensional datasets, they are less labor-intensive when it comes to feature engineering. This is due to the fact that feature-engineering is automatically carried out during the training process, which as a result makes them an ideal approach among researchers. DL-based models applications in the ML research community are extremely vast, some of which include image processing, Natural Language Processing (NLP), anomaly detection and fraud detection. However, the downsides of DL-based algorithms may bring about major hurdles that may prevent the ML models from learning efficiently \cite{8}. In ML projects, the interpretability of algorithms plays a significant part in the quality assurance process of the project. This as a result calls for ML algorithms that can be more easily analyzed. Although DL-based algorithms have shown a great deal of applicability in regards to pattern recognition tasks that are nonlinear, their behavior is still considered extremely difficult to interpret \cite{9}. Hence, the interest among researchers in regards to ML algorithms that can be both interpretable and capable of solving non-linear problems has increased. Moreover, complex DL algorithms suffer from another flaw that may prevent the deployment of ML algorithms when it comes to computing resources. RAM usage in complex DL algorithms is exponential and in some cases, may even cause the server to crash during model training. Therefore, alternative approaches are considered a necessity when abundance of computer RAM is not at our disposal. Tsetlin Machine (TM) is an exceptional ML algorithm that has proven to have near accuracy and in some cases better accuracy compared to DL algorithms in pattern recognition tasks \cite{10}. Berge et al. for instance implemented a TM-based model for text-categorization that yielded more accurate results compared to other DL-based methods \cite{11}. One of the variants of TM that can result in equal or even better accuracy is the Convolutional Tsetlin Machine (CTM) \cite{12}. The proposed method by Tunheim et al. in \cite{13} implemented an image-classification model that yields promising results compared to DL-based models. Furthermore, the performance of their model in comparison is more efficient and less intensive.

In this paper we propose an IDS called CTMBIDS (Convolutional Tsetlin Machine Based Intrusion Detection System) in the SDN environment for Distributed Denial of Service (DDoS) attacks. The key advantages of our approach compared to other complex implementations of IDSs are the interpretability of the CTM algorithm in our model as well as less memory usage that result in efficient learning and maintaining similar or better accuracy. Additionally, the SDN architecture functions as the fundamental foundation of our proposed model. Therefore, compared to traditional networks, the implementation of the proposed model can be easily done in an SDN controller. Since SDN networks have grown substantially in recent years \cite{14}, the dire need for SDN-based datasets is considered a necessity. Most IDSs that are ML-driven are trained on datasets whose data is generated in traditional networks \cite{15}. Consequently, training ML-driven IDSs on datasets that are generated in an SDN environment is still an open challenge \cite{16}. In addition to proposing CTMBIDS, we also generate three different datasets whose data stem from three distinct SDN topologies in the network. In order to have a better understanding of the CTMBIDS approach, we also provide performance evaluations compared to other common ML-based and DL-based algorithms. In summary, the main contributions of our work are as follows:

\begin{itemize}
  \item We propose a computationally less intensive IDS called CTMBIDS, which consumes much less memory and maintains relatively better accuracy in most datasets compared to DL-based IDSs.
  \item We generate three different datasets that are based on the SDN environment in order to provide our model with real SDN data so that the training of the CTMBIDS model can be as close to real scenarios as possible. We also discuss the data-generation algorithm that we used in the SDN controller in order to understand the process of obtaining the data.
  \item We also use the KDDCup99 dataset to compare the quality of our generated datasets with a well-known benchmark in the research community.
  \item In this paper, we also provide performance evaluation to compare the accuracy of our model with well-known ML-based and DL-based algorithms. These algorithms include: K-nearest Neighbor (KNN), Logistic Regression (LR), Support Vector Machine (SVM), Random Forest, Naive Bayes, Tsetlin Machine and Convolutional Neural Networks-Long Short Term Memory (CNN-LSTM).

\end{itemize}

The remaining sections of the paper are structured as follows: Section 2 discusses the necessary topics such as SDN architecture, the TM and CTM algorithm. After that, the third section will discuss the related works that proposed similar methods for implementing IDSs in a network, whether a traditional or an SDN network. Section 4 will provide a detailed explanation of our proposed method. Section 5 will talk about the empirical results in detail. Section 5 also provides a discussion of the proposed method. Finally, in section 6 a comprehensive conclusion and future work that the authors may pursue will be provided.

\section{Preliminaries}\label{sec2}
In this section, we provide a comprehensive background for SDN architecture, TM architecture, CTM architecture, and the classifiers that are applied in our work for the purpose of comparison and model evaluation. 

\subsection{SDN architecture}
Most of the existing traditional networks suffer from various shortcomings. These flaws include scalability, cost, security and manageability. Network devices are extremely costly and demand high-maintenance \cite{17}, which as a result prevents networks from expanding. This problem becomes even more difficult when vendor-based devices are added to the network, which as a result forces the network administrators to deal with different types of manual configurations. This type of arduous network management is prone to human error and may ultimately lead to unwanted security loopholes that can cause irreparable damage to the network \cite{18}.

\begin{figure}[h]
\centering
\includegraphics[width=0.7\textwidth]{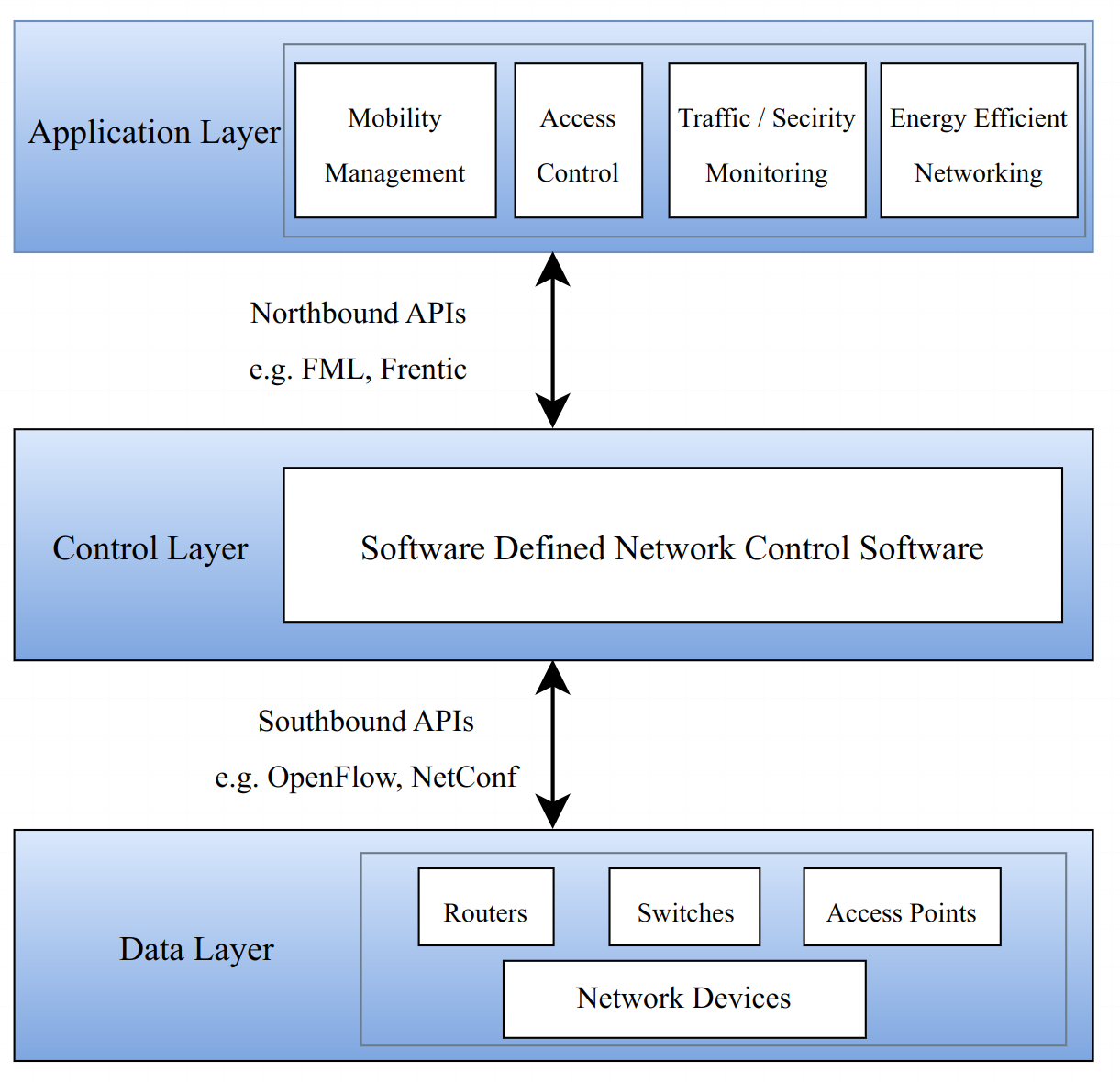}
\caption{SDN architecture and decoupling of network layers \cite{20}.}\label{fig1}
\end{figure}

SDNs on the other hand provide a flexible architecture that eliminates the shortcomings of traditional networks via decoupling the data layer and the control layer, allowing the network to be substantially cheaper and yet more scalable and more manageable in a vendor-less environment. This as a result gives room to the research community to more easily implement innovative approaches as softwares in the SDN controller \cite{19}. As figure \ref{fig1} shows, after decoupling the data and control layer, main network functions can be easily implemented as applications or softwares in the application layer. The three layers are in communication with each other via Application Programming Interface (API). Therefore, applications and network devices have no direct communication with each other. 

Although SDN's scalability brings about numerous advantages, the downside of security threats are still regarded as one of the most important parts of an SDN architecture that needs to be addressed. The work of Chica et al. in \cite{21} covered the attack surface of each layer in SDN architecture, and as it can be perceived, maintaining the reliability and security of SDN is of high priority. The focus of this paper is on DDoS intrusions in the SDN environment, hence the implementation of a novel Intrusion Detection System (IDS) for DDoS attack with lowest memory consumption and highest accuracy. 

\subsection{DoS and DDoS Attack in SDN}
\begin{figure}[h]
\centering
\includegraphics[width=0.7\textwidth]{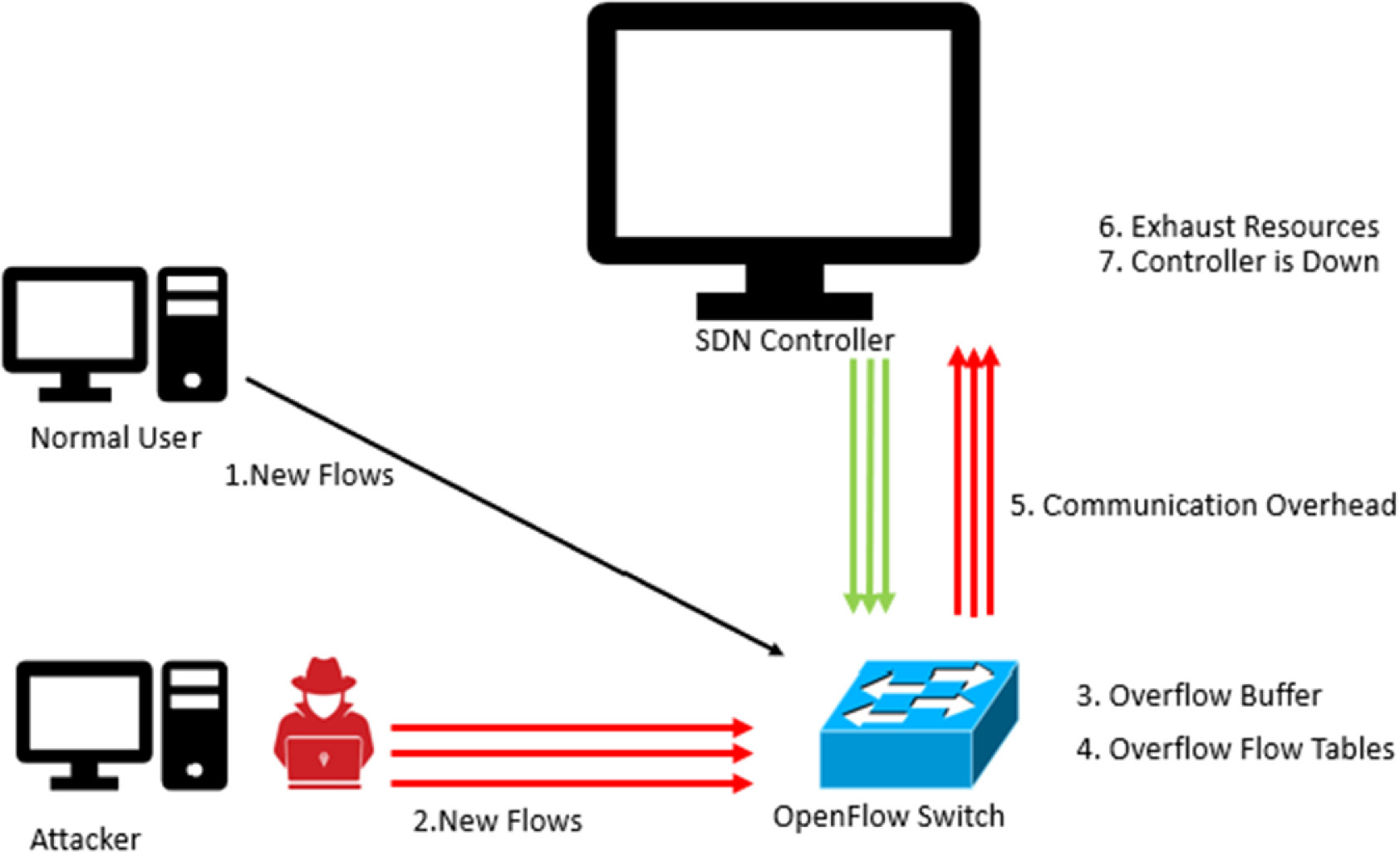}
\caption{DDoS attack in the SDN environment \cite{22}.}\label{fig2}
\end{figure}

DDoS and DoS attacks are among the most harmful types of network intrusions that can cripple the entire network via resource exhaustion or service incapacitation. Since services in the SDN environment are mounted on SDN controllers in the form of virtual applications, SDN controllers become the primary target for network attackers. In this way, the legitimate clients will be deprived of network services and are not able to communicate with network applications. As it is demonstrated in figure \ref{fig2}, the attack in the SDN environment can generate flows in an SDN environment that uses spoofed IP addresses using a single device (DoS) or attackers who do the same exploiting multiple devices (DDoS). Normally, requests made from spoofed IP addresses do not exist in the flow table of the OpenFlow switch to be matched. This, as a result, ends up with a table miss condition that ultimately forces the switch to bombard the controller with flow messages that exhaust the bandwidth, CPU and memory in the data and the control layer \cite{22}. 

Bearing the above mentioned situation and the architecture of the SDN in mind, the most optimal solution is the deployment of an anomaly-based IDS that can act as an application in the application layer of the SDN environment. In this way, DDoS and DoS attack detection can be carried out more efficiently in comparison to traditional networks. There are publicly available datasets that can be utilized in order to train a state-of-the-art IDS. In the next section we will compare and discuss these datasets. 

\subsection{DDoS Datasets}
Depending on the type of IDS that needs to be trained, datasets can play a significant role in the efficiency of the IDS. The quality of the IDS in the real-world scenarios can drop drastically if the quality of the dataset features do not meet the required standard. Table \ref{tab1} discusses the main properties of 8 most important datasets that can be used to train an IDS. The testbed for all the datasets are traditional networks except for the InSDN dataset, which was curated with 46 features in 2016. Some of these datasets have restricted access while most of them are available for training an IDS. Moreover, KDDCup99 and DARPA DDoS datasets are regarded as legacy datasets. KDDCup99 is one of the most important datasets that have been used as a reference so that other datasets can have a benchmark to be compared with. Although the dataset suffers from imbalanced data and is created in a simulated environment, its features are suited to train an IDS with. Later on, the dataset was improved under the title of NSL-KDD in order to cover the drawbacks of the KDDCup99 dataset. As it can be seen from table \ref{tab1}, the number datasets in the SDN environment are heavily outnumbered against the datasets in traditional networks. Therefore, it becomes imperative to have high-quality datasets for the SDN environment. In section 3, we will discuss our proposed generated datasets and the classification mechanism for implementing an IDS against DDoS attacks.
\begin{table}[h]
\caption{Comparison of network datasets.}\label{tab1}
\begin{tabular}{@{}p{2.29cm}p{1.49cm}p{2.2cm}lp{1.4cm}lp{1cm}@{}}
\toprule
Dataset & Type of Testbed & Type of DDoS Attack & Availability & Number of Features & Year & SDN-Oriented\\
\midrule
KDDCup99 \cite{23} & Synthetic & SYN flood & Available & 41 & 1999 & No\\
NSL-KDD \cite{24} & Synthetic & Back, Land, Neptune, Process table & Available & 42 & 2009 & No \\
DARPA DDoS Attack \cite{25} & Real-World & SYN flood & Restricted & 115 & 2009 & No\\
CAIDA \cite{26} & Real-World & UDP flood & Restricted & NA & 2009 & No\\
UNSW-NB15 \cite{27} & Real-World & Not Specified & Available & 49 & 2015 & No\\
CICIDS 2017 \cite{28} & Real-World & TCP SYN Flood, TCP SYN-ACK, UDP Flood, ICMP Flood, HTTP Flood & Available & 78 & 2017 & No\\
BoT-IoT \cite{29} & Synthetic & Not Specified & Available & 116 & 2018 & No\\
InSDN \cite{30} & Synthetic & TCP-ACK Flood, UDP Flood, HTTP Flood, HTTP POST, Slowloris, TCP Flood, TCP-SYN Flood, ICMP Flood
 & Available & 41 & 2016 & Yes\\
\bottomrule
\end{tabular}

\end{table}

\subsection{Tsetlin Machine Architecture}
\begin{figure}[h]
\centering
\includegraphics[width=0.8\textwidth]{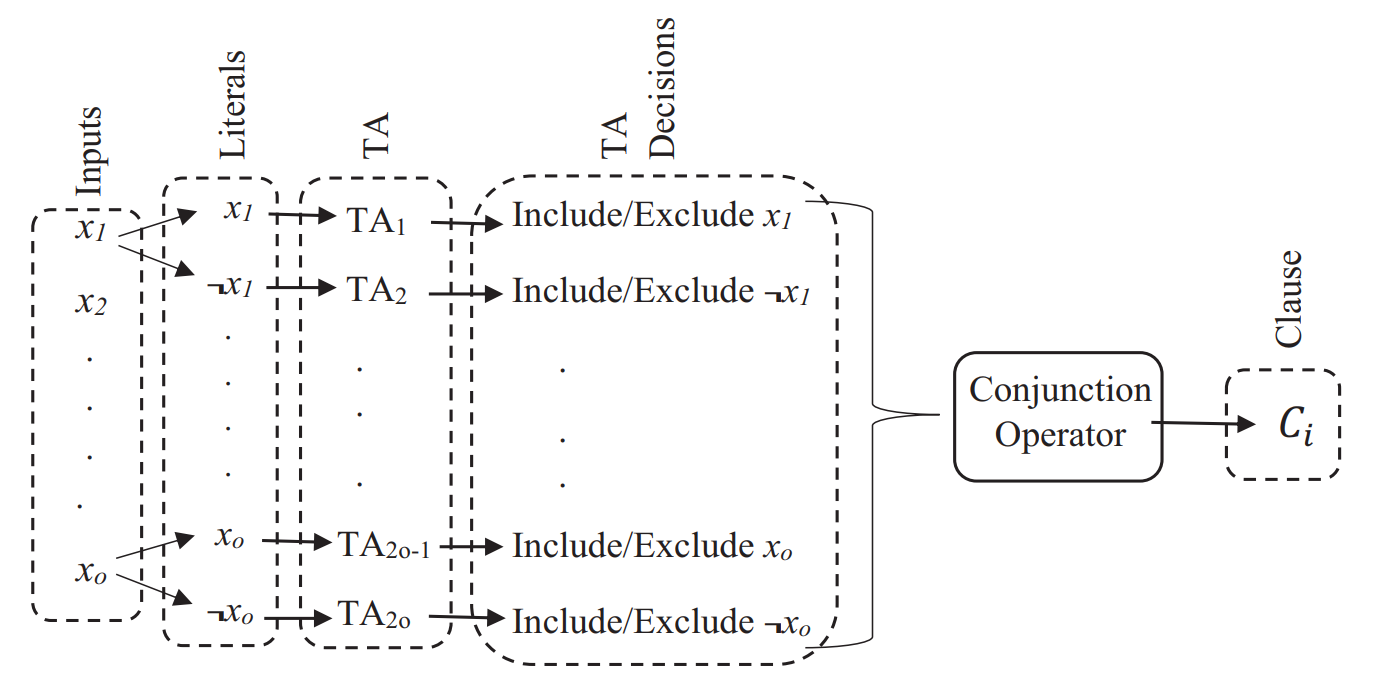}
\caption{Tsetlin automaton with 2N states and 2 Actions \cite{31}.}\label{fig3}
\end{figure}

The TM algorithm consists of many parts. However, the algorithm's core component is a two-action Tsetlin Automaton (TA) whose responsibility is to take actions and train the a TM model based on the input and the current state. This procedure follows the basic principle of reward and penalty for choosing the best action. Other than TAs, inputs of the TM algorithm also play a critical role during the training as well. A binary input and its negation form a literal in the TM algorithm. The literals are then used as inputs for automatons that decide whether to include or exclude the input during the training process. Finally, after deciding which literals to include or exclude, the TM uses an AND operator to calculate the literals in order to form a clause with either a positive or negative polarity \cite{31}. Figure \ref{fig3} demonstrates the architecture of TM from input to clause calculation.

\subsection{Convolutional Tsetlin Machine}
As the conception of CNN had a great impact on DL approaches, CTM algorithm also made room for more advanced approaches using TA in order to gain better accuracy. The CTM algorithm utilizes the same recognition procedure as the TM algorithm \cite{32}. However, there are distinct differences between the two. We will discuss them in this part to better understand the CTM algorithm. Just like the CNN algorithm, the CTM algorithm's biggest advantage mostly revolves around image processing tasks. Imagine an image size of $X + Y$, which also consists of $Z$ layers. Therefore, in comparison to the classical TM algorithm, each clause of the CTM algorithm functions as a kernel. The kernel's dimensionality is $W \times W$. Considering that each image has $Z$ layers, each clauses will ultimately have $W \times W \times Z \times 2$ literals. Another big advantage of CTM is its location-awareness functionality. Consequently, this allows the model to identify patterns and memorize their locations. The final evaluation of each CTM kernel is $B = B_X \times B_Y$. In this equation, $B_X = [\frac{X-W}{d}]+1$ and $B_Y = [\frac{Y-W}{d}]+1$, in both of which $d$ is the kernel step of the convolution.
In the end, once the convolution process is finished, the CTM algorithm produces $B$ values per image, which is on the contrary to the classic TM algorithm that outputs only a single value. $C_j^{b,+}$ is denoted as the output of a positive clause $j$ on kernel $b$. In order to convert multiple outputs $C_j^{1,+}$, ... , $C_j^{B,+}$ of clause $j$ with positive polarity into one single output, which we denote as $C_j^+$, we need to utilize logical OR for all the outputs as shown in the equation (\ref{eq1}) below:

\begin{equation}
C_j^+ = \bigvee_{b=1}^B C_j^{b,+}\label{eq1}
\end{equation}

\subsection{CTM Classification}
The classification in the CTM algorithm happens exactly the same way as the TM algorithm. It is based on two-class data, i.e., the subpatterns in each class should be detected. Therefore, the classes are divided into two distinct clauses; clauses with even indices whose role is to detect subpatterns associated with output 0 are given negative polarity $(C_j^-)$, and clauses with odd indices whose role is to detect subpatterns associated with output 1 are given positive polarity $(C_j^+)$. As soon as a subpattern is detected, the CTM algorithm moves forward with a voting procedure among all of the clauses based on their polarity. The voting procedure is the equation (\ref{eq2}):

\begin{equation}
v = \sum_j C_j^+ - \sum_j C_j^- \label{eq2}
\end{equation}
which consists of a summation of all of the clauses with negative polarity subtracted from summation of all of the clauses with positive polarity. Finally, the classification output $\widehat{y}$ will be according to the condition below, which interestingly resembles the functionality of ReLU function:

\begin{equation}
\widehat{y} = \begin{cases}
1 & \text{if } v \geq 0 \\
0 & \text{if } v < 0
\end{cases}
\label{eq3}
\end{equation}

\subsection{CTM Computational Complexity}
The computational complexity of the CTM algorithm increases linearly as the number of clauses $m$ and kernel $B$ increases. Yet, the computations can be parallelized due to the decentralized nature of the CTM algorithm. This complexity is somewhat different in the TM algorithm since the absence of a kernel in the nature of TM algorithm makes a distinction. Therefore the big O for the CTM algorithm is $O(m  B)$, where, as mentioned above, $m$ represents clauses and $B$ the kernels in the algorithm \cite{12}.

\subsection{CTM hyperparameters}
After thorough examination of the CTM algorithm, we can discuss the hyperparameters of this algorithm. Table \ref{tab2} provides a description of hyperparameters that can be set in the CTM algorithm. As it is clearly explained, for each $N$ clause, we need to double the states. Also, the learning sensitivity is the sensitivity of the algorithm during training to change its states in regards to its input. In other words, it is the equivalent of learning rate in deep learning models. We will discuss the hyperparameters of our proposed model CTMBIDS in the Section 4.

\begin{table}[h]
\caption{Hyperparameters in the CTM algorithm.}\label{tab2}%
\begin{tabular}{@{}ll@{}}
\toprule
Hyperparameter & Description \\
\midrule
$N$ & Number of inputs\\
$m$ & Number of clauses\\
$2N$ & Number of automaton states\\
$T$ & Feedback threshold\\
$s$ & Learning sensitivity\\
$k$ & Kernel size \\
$w$ & Initial weight of each clause\\
\botrule
\end{tabular}

\end{table}

\subsection{Clause Feedback Activation Function}
CTM algorithm utilizes two types of feedback, both of which take advantage of clause feedback activation function. The use case of this function becomes prominent when a certain type of feedback can become problematic as soon as its frequency is left unchecked. Therefore, a new function is required to manipulate the frequency of feedback for a particular pattern, so that high-frequency and unnecessary feedback can be prevented in the model. Hence, to enhance the allocation of sparse pattern representation resources offered by the clauses, equation (\ref{eq4}) and (\ref{eq5}) are utilized for feedback type I and type II respectively, in order to decrease the frequency of each type of feedback as the number of clauses voting from that pattern approaches a threshold value $T$.

\begin{equation}
\frac{T - max(-T, min(T, \sum_{I = 1}^m C_j)}{2T}
\label{eq4}
\end{equation}

\begin{equation}
\frac{T + max(-T, min(T, \sum_{I = 1}^m C_j)}{2T}
\label{eq5}
\end{equation}
\\
As it can be perceived from the probability of activation functions, the probability of activation decreases as the number of votes approaches the threshold $T$ and ultimately as soon as $T$ is reached, the probability equals $0$. This indicates that Type I feedback will not be activated when enough clauses produce the correct number of votes, causing the affected clauses to become "frozen" as TAs will no longer change their state. Thus, other clauses are freed to seek other sub-patterns since the "frozen" pattern is no longer seeked by the TAs. This same reasoning applies to Type II feedback via adopting the clause feedback activation function, which in turn makes room for more effective resource allocation. 

\section{Related Works}
In this section, we will investigate various famous works that are ML- and DL-based. We will also discuss their results and the algorithm that they utilized to achieve the desired output. 
Tan et al. \cite{33} proposed an IDS within wireless sensor networks that utilizes Random Forest algorithm combined with Synthetic Minority Oversampling Technique (SMOTE). Their method utilizes SMOTE for increasing the accuracy and oversampling the data. After the implementation of oversampling using SMOTE, the training set is constructed again in order to balance the imbalanced class of the data. Thereafter, they use Random Forest for training on the balanced data. Their method was implemented on KDDCup99 dataset \cite{34}, whose end result after the use of SMOTE demonstrated that precision and accuracy on imbalanced classes of attacks in the dataset were improved compared to the precision and accuracy without using SMOTE. SMOTE is not the only methodology that can deal with imbalanced data. Wazirali in \cite{34} introduced a semi-supervised IDS whose performance was improved using hyper parameter tuning on KNN algorithm. This approach utilized 5-fold cross validation strategy for model validation until the overall k is achieved. The model also utilizes principal Component Analysis (PCA) in order to select the most relevant features in the data. The proposed method achieves the highest accuracy compared to other proposed models.

\begin{sidewaystable}
\caption{Comparison of methods for intrusion detection in software-defined networks.} \label{tab3}
\begin{tabular}{p{2.6cm}lp{1.5cm}p{3cm}lp{2.5cm}p{2.5cm}p{2.5cm}}
\toprule
Methods & Year & Approach & Datasets & SDN based & Big O Notation & Advantages & Disadvantages \\
\midrule
Besharati et al. \cite{38} & 2019 & Logistic Regression & NSL-KDD & No & $O(m \times n \times k)$, the number of training examples ($m$), features ($n$) and optimization iterations ($k$). & New approach for feature selection, Improvement in accuracy. & Lack of details in practical implementation. \\
Wazirali \cite{34} & 2020 & KNN & NSL-KDD & No & $O(d \times n \times \log(k))$, the dimensionality ($d$), the number of training examples ($n$) and the number of neighbors ($k$). & Improvement of KNN detection rate, computationally efficient on resource constrained devices. & Requires a large amount of data for high performance, Sensitive to the choice of hyperparameters. \\
Anton et al. \cite{35} & 2019 & SVM & Modbus-based gas pipeline control traffic, OPC UA based batch processing traffic. & No & $O(m^2 \times n)$, the number of training examples ($m$) and features ($n$). & Capable of detecting a variety of industrial protocols, including Modbus and OPC UA, Capable of handling missing data. & Vulnerability to false positives.\\
Tan et al. \cite{33} & 2019 & Random Forest & KDDCup99 & No & $O(m \times n \times log(n))$, the number of trees ($m$), features ($n$) and $\log(n)$ due to sorting operations. & Capable of handling imbalance data. & Vulnerability to false positives and dependency on hyperparameter values.\\
Wisanwanichthan and Thammawichai \cite{36} & 2021 & Naive Bayes & NSL-KDD & No & $O(m \times n)$, the number of training examples ($m$) and features ($n$). & High detection rates and low false alarm rates, Robust to noise and outliers. & Complex model which lacks interpretability, Sensitivity towards the parameters.\\
Abeyrathna et al. \cite{40} & 2020 & Tsetlin Machine & KDDCup99 & No & $O(m)$,the number of clauses ($m$). & Interpretability of the model, Competitive performance compared to recent IDS. & Lack of details in preprocessing step.\\
Abdallah et al. \cite{39} & 2021 & CNN-LSTM &NSL-KDD & Yes & $O(L \times n)$, processing data through layers ($L$) with an input size of ($N$). & Capability in capturing temporal features, Improved performance in terms of generalization. & Demanding in computations, Complex compared to other IDSs.\\
\bottomrule
\end{tabular}
\end{sidewaystable}

Anton et al. \cite{35} proposed an IDS that utilizes a hybrid model taking advantage of both SVM and Random Forest algorithms in an industrial environment. Their model utilizes Random Forest Importance Score for feature selection as well as SVM for the classification part of the model. In addition to using Random Forest for feature selection, their method also utilized a full-feature dataset, which in the end resulted in a very promising accuracy compared to the accuracy of the model when features were selected. The result also showed that, in spite of relatively good accuracy, the model training was extremely slow when using SVM algorithm. Naive Bayes is another approach that can yield relatively good results in classification tasks. Wisanwanichthan and Thammawichai \cite{36} utilized Naive Bayes and SVM as a hybrid model on the KDDCup dataset. Their method obtains the highest accuracy among other approaches on two types of attacks in the KDDCup dataset. The proposed model utilizes Intersectional Correlated Feature Selected (ICFS) in order to drop the irrelevant features. However, their proposed method suffers from bias toward frequent attacks, which signifies that the proposed model underestimates the attacks that occur less frequently.

Although traditional machine learning models are applicable approaches, gaining high accuracy while dealing with large datasets and preprocessing the data can be frustrating \cite{37}. Therefore, it becomes a necessity for IDSs to harness the power of DL-based models. However, the one biggest disadvantage of DL models is memory consumption that may be considered an obstacle when hardware resources are limited. Besharati et al. proposed a host-based IDS that utilizes Logistic Regression inside VMs in cloud environments \cite{38}. Their framework takes advantage of the NSL-KDD dataset. However, their significant contribution is using the logistic regression algorithm for feature selection. Their work yields a \%97.51 accuracy in the cloudsim environment. Their one big disadvantage, according to the authors, is the additional computational complexity that makes the proposed model somewhat more complex. Abdallah et al. proposed a hybrid model that uses a combination CNN and Long Short-Term Memory (LSTM) algorithms. This approach made an effort to utilize the capabilities of both algorithms, that is, the spatial feature extraction capability of CNN algorithm and the temporal feature extraction capability of LSTM algorithm. The evaluation of the proposed model was carried out on three datasets: CIC-IDS 2017, UNSW-NB15, and WSN-DS \cite{39}. The criteria for evaluating the model's performance were accuracy, precision, detection rate, F1-score, and False Alarm Rate (FAR). Apart from deep learning approaches, TM algorithm can also be considered another alternative when maintaining the high accuracy of deep learning models is in play. In fact, the TM algorithm not only can maintain the relatively high accuracy, but it can also maintain the interpretability of the model. Abeyrathna et al. implemented an IDS using the TM algorithm that was trained on the KDDCup99 dataset. Their model outperformed numerous other algorithms including SVM, Decision Tree and Random Forests \cite{40}.

Finally, It is also worth mentioning that, to the best of our knowledge, Mohsin et al.'s study in \cite{41} employed a similarly generated dataset to address the security concerns posed by DDoS attacks. In spite of claiming to generate a new dataset, their work has the fundamental flaw of not providing any technical details of their data generation algorithm. Furthermore, their method fails to demonstrate the necessary details in regards to the generated data such as the type of generated features and the necessary preprocessing steps for cleaning the generated dataset. 

Considering the above discussed methods, the works are introduced to address the threat of network intrusions either in SDN environments or in traditional networks using state-of-the-art approaches. We selected these works and presented a summary of them in table \ref{tab3} in order to later on compare the proposed CTMBIDS method with these approaches. As it is evident, each work has a unique approach, which allows us to have a comprehensive comparison. Moreover, the advantages and disadvantages of each work make room for a more in-depth understanding of this comparison so that it can be used as a benchmark for future works. As it can be seen, each work in the table is provided with its advantages and disadvantages as well as its algorithm, the dataset that was used for the proposed IDS and finally whether or not the work is implemented in an SDN environment.

\section{Proposed Method}
\begin{figure}[h]
\centering
\includegraphics[width=1.0\textwidth]{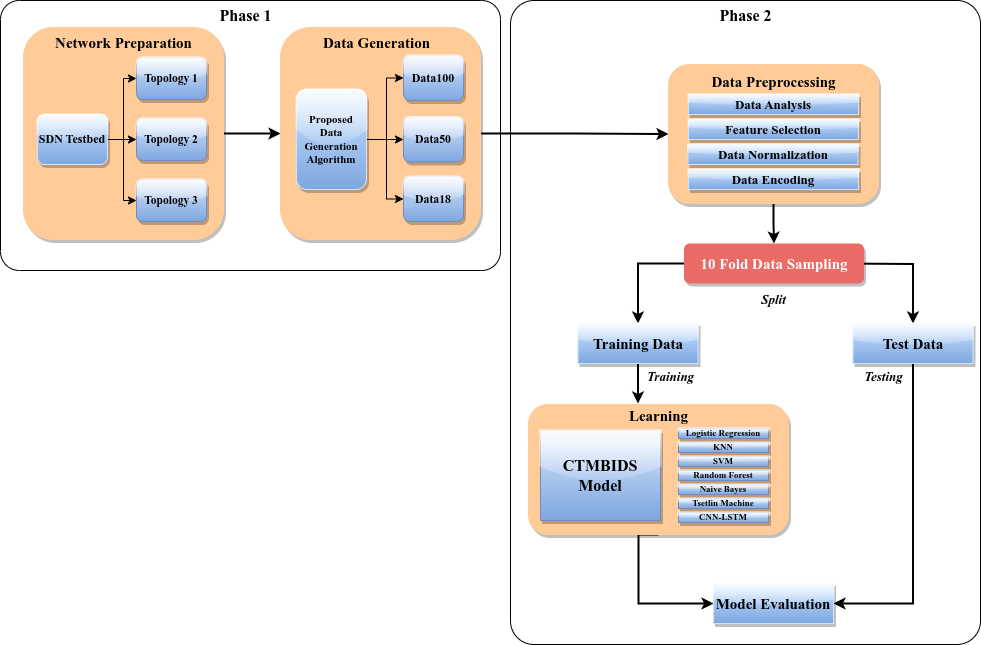}
\caption{Workflow of the proposed CTMBIDS method.}\label{fig4}
\end{figure}

Our proposed method consists of 2 phases. Phase 1 consists of preparation of the SDN testbed for data generation. Phase 2 consists of data preprocessing and model training. In this phase, we propose a model called CTMBIDS for detecting DDoS attacks in an SDN environment. Figure \ref{fig4} shows the workflow demonstrating the processes of each phase in detail. Hence, in this section, we go over each phase and its details in order to discuss what each phase consists of.

\subsection{Phase 1: Network Topologies}
As we discussed earlier, one of the main contributions of our work is data generation so that the trained models can learn on real world SDN data. For this reason, the first step of our model starts with three main network topologies in an SDN environment. These topologies differ in size of their hosts and their network devices. This is to make sure that the generated data is not biased towards a specific network topology. Moreover, three topologies are an indication that there are three different datasets for model training, which can be extremely useful for evaluating the model's performance with other existing models and datasets. Table \ref{tab4} shows the details of our three SDN network topologies. As it can be seen, each topology consists of a number of Virtual Machines (VMs) that are connected to the SDN Open vSwitch. For example in the third topology, there are 10 SDN switches and for every switch, there are 10 VMs that are connected to it.

\begin{figure}[h]
\centering
\includegraphics[width=0.7\textwidth]{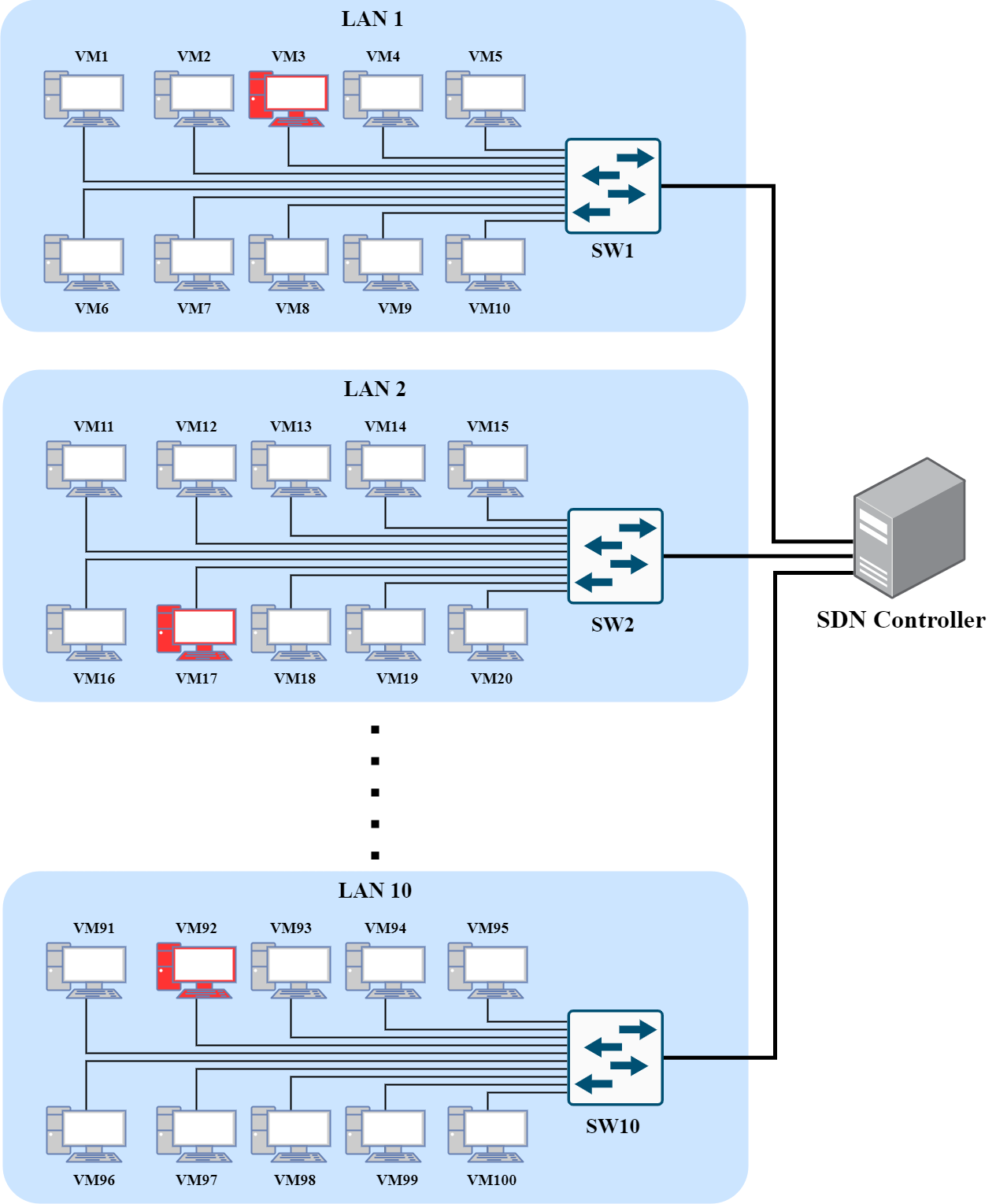}
\caption{SDN testbed for the data generation procedure in topology 3.}\label{fig5}
\end{figure}

\begin{table}[h]
\caption{Details of SDN Network Topologies.}\label{tab4}%
\begin{tabular}{lllll}
\toprule
Topology & Total VMs & VMs in a single LAN & SDN Switches In LAN & SDN controller \\
\midrule
Topology 1 & 18 & 3 & 6 & Ryu Controller \\
Topology 2 & 50 & 5 & 10 & Ryu Controller \\
Topology 3 & 100 & 10 & 10 & Ryu Controller \\
\botrule
\end{tabular}
\end{table}
Figure \ref{fig5} shows the SDN testbed for topology 3, which consists of 10 switches and 100 VMs or nodes. In each Local Area Network (LAN), we have 9 VMs whose network traffic is normal across the entire network while 1 host in each LAN is randomly chosen to generate malicious traffic. We will discuss the data generation part of the method more in the next section. The network controller for all topologies is Ryu Controller. The reason for choosing this controller is its compatibility with the Python programming language, which is the same programming language that we used throughout the entire steps of the proposed CTMBIDS model.

As for the data layer of all the topologies, we utilized Open vSwitch alongside the Mininet simulator for creating all the nodes in the data layer. Other alternatives for creating the data layer existed but due to its compatibility with Ryu controller, we decided to use Open vSwitch. It is worth noting that version 1.3 of OpenFlow was used for all the Open vSwitches in all the topologies that we mentioned. 

\subsection{Phase 1: Data Generation Algorithm}
\begin{figure}[h]
\centering
\includegraphics[width=0.6\textwidth]{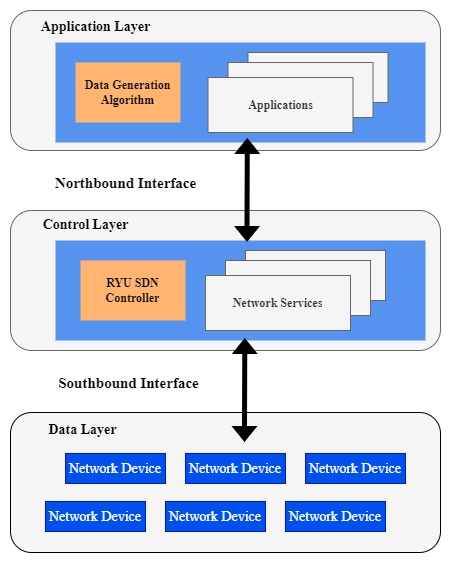}
\caption{The placement of the DataGenerationApp algorithm in the proposed SDN architecture.}\label{fig6}
\end{figure}

After creating three different topologies, we used the proposed data generation algorithm to generate three distinct datasets. As their name suggests, Data18 is created based on the topology that has 18 hosts, Data50 is generated based on the topology that has 50 hosts and Data100 is generated based on the topology that has 100 hosts, respectively. The DataGenerationApp algorithm functions as a software in the application layer of the SDN architecture. Figure \ref{fig6} demonstrates the placement of the algorithm in the proposed SDN architecture. As it can be seen in figure \ref{fig6} , the SDN network devices are located in the data layer. They interact with the Ryu controller through the southbound interface. This interface is responsible for relaying the messages between the control layer and the data layer. In other words, the Ryu controller utilizes the southbound interface in order to configure, manage and collect data from the existing devices. The application layer on the other hand, utilizes northbound interface to retrieve information from the control layer in the form of events. This process uses protocols that take advantage of SDN controller event handlers. As a result all network devices are exposed to applications via the controller so that the end result can be information retrieval and device management in the SDN environment.

Algorithm \ref{algo1} explains how the proposed algorithm called DataGenerationApp functions in the application layer in order to perform data generation across the entire network. At first, all the available datapaths in the network are stored in a python dictionary, in which the datapath id is its key and the datapath information is its value. Afterwards, using an infinite loop, the DataGenerationApp asynchronously and continuously monitors the OpenFlow switches through a request that is generated from the Ryu controller event handler in order to request statistics from OpenFlow Switches. The controller acts as an intermediary between network switches and the DataGenerationApp. This process also includes a 10-second inactivity gap after each event handler request simply due to the fact that we wanted to make sure we write all the requested information. The main function FlowStatsReplyHandler receives a reply from the switch in order to get the flow statistics message. This function's input is the statistics message that was sent from the switch to the controller and function's main purpose is to return the statistics of the specified features. All the openflow flow statistic reply and request from the input message is implemented using Ryu event handler methods since the Ryu controller is the intermediary between switches and the DataGenerationApp. This indicates that the FlowStatsReplyHandler function acts as a "parser" that analyzes the flow statistics reply and then writes the statistics to a CSV file. Once the statistics of the message is parsed, the generated data is annotated as 0 if the parsed statistics are of normal traffic. Otherwise, the statistics are parsed as malicious traffic, which means that data should be annotated as 1. 

\begin{algorithm}
\caption{DataGenerationApp}\label{algo1}
\begin{algorithmic}[1]
\Require datapath $d$ of OpenFlow switch, message $msg$
\Ensure  CSV file
\While{True}
	\State Initialize a random datapath
	\For{\texttt{each datapath $d$}}
		\State\text{Send flow statistics request to switch $sw$}
	\EndFor
	\State Sleep for 10 seconds
	\State Get flow statistics message $msg$ from switch $sw$
	\State Extract timestamp, datapath ID, flow ID, IP source, IP destination, Source Port, Destination Port, IP protocol, ICMP Code, ICMP Type, Duration in seconds, duration in nanoseconds, idle timeout, hard timeout, flags, packet count, and byte count from $msg$

	\If{$msg$ is Normal Traffic}
		\State Annotate data as $0$
	\Else
        \State Annotate data as $1$
	\EndIf
	\If{CSV file is not open}
		\State Open CSV file for writing
		\State Write header row to CSV file
	\EndIf
\EndWhile
\State Save CSV file
\State Close CSV file
\end{algorithmic}
\end{algorithm}

\subsection{Phase 1: Data Generation Tools}
The creation of the datasets consisted of two major parts. The first part was gathering normal traffic across the network via the DataGenerationApp. This part relied only on the legitimate devices in the network. The second part however, relied on the randomly chosen hosts in each LAN. As it can be seen from figure \ref{fig5}, one host is randomly chosen inside each LAN to generate malicious traffic. In our proposed method, these malicious data consist of DDoS attacks. Table \ref{tab5} shows the tools we used to generate the attack traffic in our dataset. We used Micro Core Linux as the main OS for each device in the network. The reason for choosing this lightweight OS is simply due to the fact that the high number of hosts in network topologies require so much memory. Therefore, choosing an OS such as Kali Linux with too many redundant tools that may not be used during the data generation process as far as DDoS attacks are concerned. The only difference in DoS and DDoS attacks are the number of attacking hosts in the network that constitutes different types of DoS attack.
\begin{table}[h]
\caption{Description of tools that were used for malicious traffic generation.}\label{tab5}%
\begin{tabular}{lp{3.95cm}ll}
\toprule
Attack Class & Attack Type & Attack Tools & Attacker Machine \\
\midrule

DoS Attack & ICMP Flood, UDP Flood, TCP-Ack Flood, TCP-Flood, SMURF Attack & Hping3, Nping, Metasploit & Micro Core Linux \\

DDoS Attack & ICMP Flood, UDP Flood, TCP-Ack Flood, TCP-Flood, SMURF Attack & Hping3, Nping, Metasploit & Micro Core Linux\\
\botrule
\end{tabular}
\end{table}

\subsection{Phase 1: Dataset Description}
As mentioned earlier, Data18, Data50 and Data100 were the three datasets that we created. Table \ref{tab6} shows the detailed information about the features of the three main datasets. As table \ref{tab6} demonstrates, we created 21 features, which include 7 categorical and 14 numerical features. Later in the next section, we will discuss the feature selection methods that we used to discuss what features were used for the CTMBIDS model training.

\begin{table}[h]
\caption{Description of all features in the three generated datasets.}\label{tab6}%
\begin{tabular}{llll}
\toprule
id & Feature & Description & Data Type \\
\midrule
f1 & timestamp & OpenFlow timestamp during traffic generation & numerical\\
f2 & datapath\_id & The unique ID of the OpenFlow switch & categorical \\
f3 & flow\_id & A unique identifier for the flow & numerical \\
f4 & ip\_src & The IP address of the source of the flow & categorical \\
f5 & ip\_dst & The IP address of the destination of the flow & categorical \\
f6 & tp\_src & The transport port of source of the flow & numerical \\
f7 & tp\_dst & The transport port of destination of the flow & numerical \\
f8 & ip\_protocol & The IP protocol of the flow & categorical \\
f9 & icmp\_code & The ICMP code (if the protocol is ICMP) & categorical \\
f10 & icmp\_type & The ICMP type (if the protocol is v4 or v6) & categorical \\
f11 & flow\_duration\_sec & The duration of the flow in seconds & numerical \\
f12 & flow\_duration\_nsec & The duration of the flow in nanoseconds & numerical \\
f13 & idle\_timeout & The idle timeout of the flow in seconds & numerical \\
f14 & hard\_timeout & The hard timeout of the flow in seconds & numerical \\
f15 & flags & The flags of the flow & categorical \\
f16 & packet\_count & The number of packets in the flow & numerical \\
f17 & byte\_count & The packet bytes in the flow & numerical \\
f18 & packet\_count\_per\_second & The packet count per second in the flow & numerical \\
f19 & packet\_count\_per\_nsecond & The packet count per nanosecond in the flow & numerical \\
f20 & byte\_count\_per\_second & The packet bytes per second in the flow & numerical \\
f21 & byte\_count\_per\_nsecond & The packet bytes per nanosecond in the flow & numerical \\
\botrule
\end{tabular}
\end{table}

All the traffic generation processes were implemented in the Mininet simulator as well as OpenFlow and Open vSwitch. Mininet is a simulation tool that is used extremely in the research community and has gained popularity with Python developers. However, in order to make sure that the generated datasets were close to other standard datasets, we also used KDDCup99 as our reference dataset. Table \ref{tab7} shows the details of each generated dataset in terms of numbers of data and the time it took for the data generation process. 

\begin{table}[h]
\caption{Description of all features in the three generated datasets.}\label{tab7}%
\begin{tabular}{lllll}
\toprule
Dataset & Normal Traffic & Malicious Traffic & Number of Features & Data Generation Time \\
\midrule
Data18 & 906,211 & 1,123,402 & 21 & Almost 48 Hours \\
Data50 & 658,884 & 723,244 & 21 & Almost 48 Hours \\
Data100 & 2,249,229 & 2,438,701 & 21 & Almost 96 Hours\\
\botrule
\end{tabular}
\end{table}

\subsection{Phase 2: Data Preprocessing}
After generating data using three different network topologies, the three datasets were created with 21 features. However, preprocessing is an essential prerequisite before model training. Therefore, in our proposed method, the preprocessing consists of 4 different steps, which are data analysis, data normalization, feature selection and data encoding. The following sections describe each step of preprocessing and how it results in four preprocessed datasets that are ready to be used for CTMBIDS training. 

\subsection{Phase 2: Data Analysis}
\begin{figure}[h]
\centering
\includegraphics[width=0.9\textwidth]{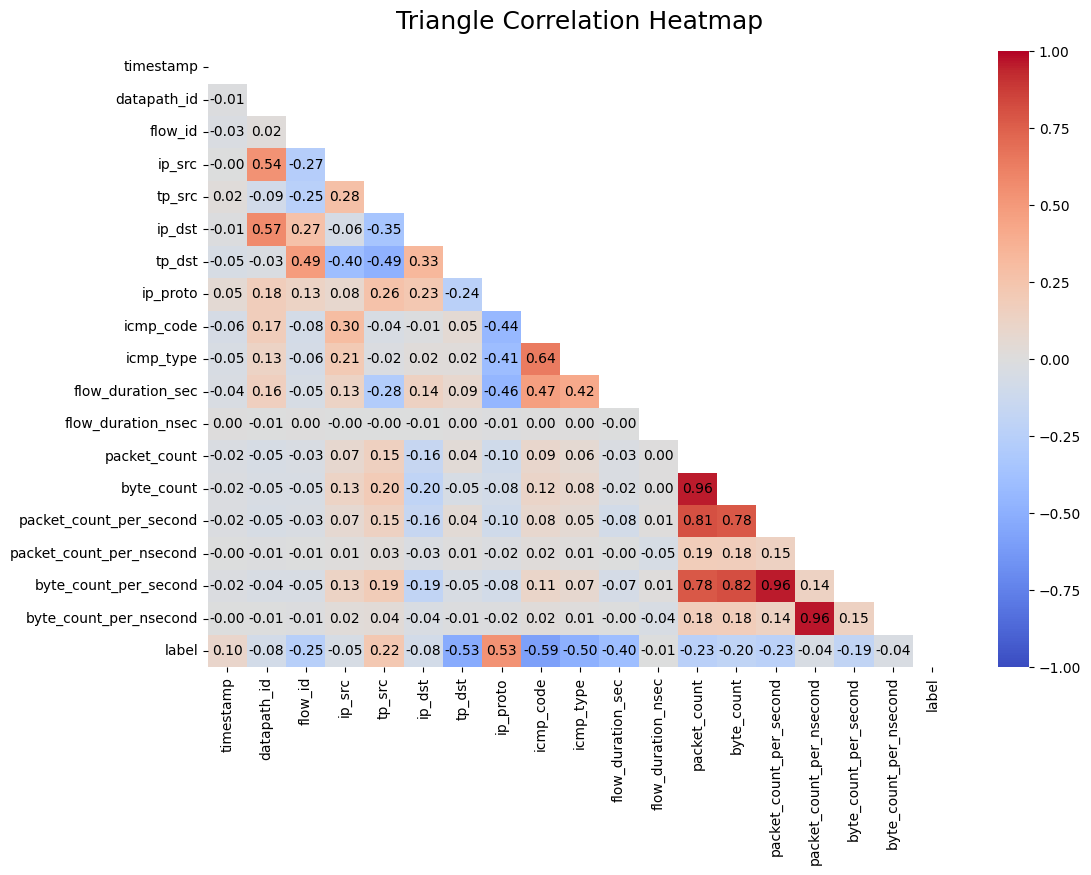}
\caption{The triangle correlation heatmap for Data100.}\label{fig7}
\end{figure}

In order to better understand the data that were created, we perform a comprehensive data analysis so that we could explore what data type each column feature contained. With the analysis of our data, not only we could better understand what type of encoding we must use later, but also how much correlation there is between one column feature and another. Figure \ref{fig7} shows the triangle correlation heatmap for one of our generated datasets, namely Data100. The correlation heatmap illustrates that the most correlated features are byte count and packet count both in terms of size and time duration in flow messages. Other dataset features that are fairly correlated with each other are source and destination port with source and destination IP addresses. 

\subsection{Phase 2: Data Normalization}

An IDS with high accuracy in classification requires data that has been normalized and preprocessed. The reason for the utilization of data normalization is because the generated as well as the KDDCup99 datasets all have numerical values that have different ranges. Thus, rescaling and normalizing the data becomes essential. In our CTMBIDS method, for rescaling and normalizing the data, we used min-max normalization for each numerical column feature according to the equation (\ref{eq6}), where $x$ is the value to be normalized,  $x_{min}$ is the smallest value in a feature column and $x_{max}$ is the largest value in a feature column:

\begin{equation}
x_{normalized} = \frac{x - x_{min}}{x_{max}-x_{min}}\label{eq6}
\end{equation}

\subsection{Phase 2: Feature Selection}
All the three generated datasets have 21 features. Therefore, reducing the number of features can be beneficial to gain more accuracy for training the CTMBIDS model. In our proposed method, we used Recursive Feature Elimination (RFE) with a Random Forest classifier using gini index \cite{42}. This algorithm recursively iterates over all the features. Once a feature is selected to be removed, the Random Forest classifier evaluates its performance on the remaining features. This process occurs iteratively until the desired number of features is reached. After implementing the RFE on both the three generated datasets and the reference KDDCup99 dataset, 16 features were chosen from each dataset. Table \ref{tab8} shows the selected features by RFE algorithm in the three generated datasets as well as in the KDDCup99 dataset. The selected features for the generated datasets in table \ref{tab8} are chosen from all the features that we demonstrated in table \ref{tab6}. The selected features for the KDDCup99 dataset are chosen from the KDDCup99 dataset.

\begin{table}[h]
\caption{The selected features for the generated and the KDDCup99 dataset using RFE algorithm.}\label{tab8}
\begin{tabular}{p{1.6cm}|*{16}{p{0.27cm}}}
\toprule
Selected Features in Generated Datasets & \multirow{4}{*}{f2} &  \multirow{4}{*}{f3} &  \multirow{4}{*}{f6} &  \multirow{4}{*}{f7} &  \multirow{4}{*}{f8} &  \multirow{4}{*}{f9} &  \multirow{4}{*}{f10} &  \multirow{4}{*}{f11} &  \multirow{4}{*}{f12} &  \multirow{4}{*}{f13} &  \multirow{4}{*}{f16} &  \multirow{4}{*}{f17} &  \multirow{4}{*}{f18} &  \multirow{4}{*}{f19} &  \multirow{4}{*}{f20} &  \multirow{4}{*}{f21} \\
\midrule
Selected Features in KDDCup99 \cite{23} & \multirow{4}{*}{f2} & \multirow{4}{*}{f3} & \multirow{4}{*}{f4} & \multirow{4}{*}{f5} & \multirow{4}{*}{f6} & \multirow{4}{*}{f10} & \multirow{4}{*}{f13} & \multirow{4}{*}{f23} & \multirow{4}{*}{f24} & \multirow{4}{*}{f29} & \multirow{4}{*}{f30} & \multirow{4}{*}{f33} & \multirow{4}{*}{f34} & \multirow{4}{*}{f35} & \multirow{4}{*}{f36} & \multirow{4}{*}{f37} \\
\bottomrule
\end{tabular}
\end{table}

Feature selection is extremely beneficial when computational resources are not abundant. However, in order to compare the quality of the generated datasets with our reference KDDCup99 dataset, and also to compare the performance of the feature selection algorithm, we will also train the CTMBIDS model with full feature datasets as well. Not only this can provide a benchmark for future reference, but also a baseline for the accuracy of the proposed CTMBIDS.

\subsection{Phase 2: Data Encoding}
Almost all the existing machine learning algorithms require an encoding scheme to prepare the data for model training. As stated above, both numerical and categorical features exist in all datasets. Therefore, it is a necessity to use the appropriate encoding scheme for each type of data type. It is worth mentioning that encoding in the proposed method was implemented after the data normalization. We used one-hot encoding to encode categorical features and thresholding method for binarizing the normalized numerical features \cite{43}. Hence, the final datasets that were used to be fed to the CTMBIDS model were all binary. This is an inseparable part of preprocessing for the TM and CTM algorithm since they work with binary data. The reason that the TM and CTM algorithms consume less memory is simply due to utilization of binary data, which is the closest language to the machine language. Finally, it is also worth mentioning that the thresholding method was only used for our proposed CTMBIDS method. Other models that we implemented in order to compare the CTMBIDS model with, took one-hot encoding and data normalization steps without utilizing thresholding method.

\subsection{Phase 2: Data Sampling}
Any machine learning model requires a training and a test set so that the desired model can learn with the training set and it can be evaluated with the test set. However, due to the novelty of our generated datasets, the CTMBIDS model as a result requires a strategy to avoid overfitting. Therefore, for the evaluation criterion of all models as well as the CTMBIDS model, we used k-fold cross validation with 10 being the k value in the proposed CTMBIDS model \cite{44}. In the end, the final average accuracy, precision and f1-score of all iterations are measured after 10-fold cross validation is completed.

\subsection{Phase 2: CTMBIDS Learning}
\begin{figure}[h]
\centering
\includegraphics[width=0.8\textwidth]{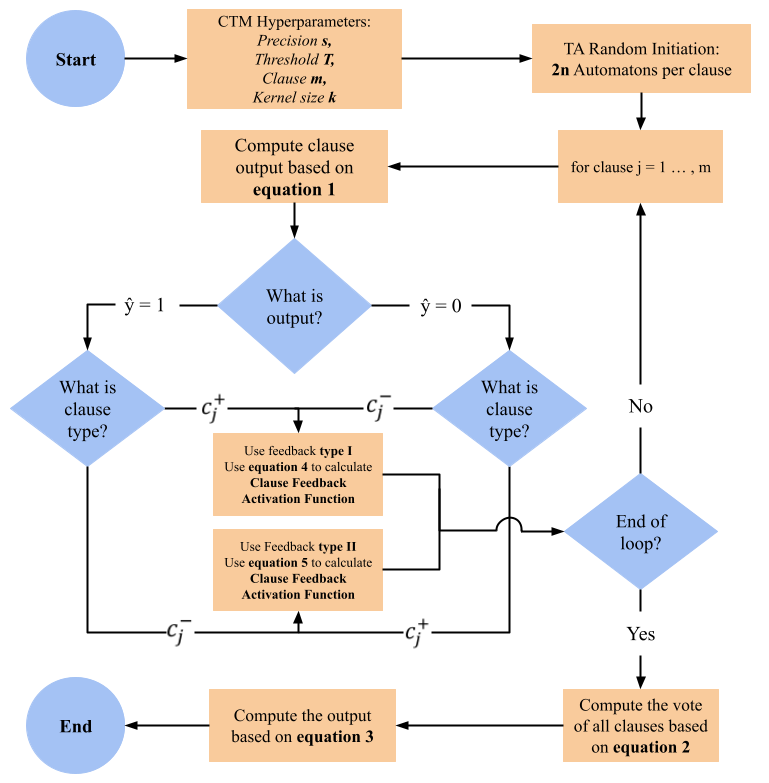}
\caption{The training workflow of the CTMBIDS model.}\label{fig8}
\end{figure}

Once the train and test sets are preprocessed and then divided into 10 folds, the training of CTMBIDS model can commence. The main classification algorithm in the CTMBIDS model is the CTM algorithm, which uses automatons in its core. Figure \ref{fig8} represents the training workflow of the CTMBIDS model. First, the hyperparameters precision $S$, threshold $T$, clause $m$ and kernel size $k$ must be set. Table \ref{tab9} shows the hyperparameters values before the training. Once the hyperparameters are set, the CTMBIDS continues to create $2n$ clauses. Afterwards, a loop iterates over every clause $m$ to calculate each clause output. The output for each clause $m$ is calculated using equation (\ref{eq1}). The next step for the CTMBIDS method depends on two things, the clause output and the clause polarity. For instance for the $j^{th}$ clause, if the clause output is 0 with negative polarity, the CTMBIDS uses feedback type I. Moreover, to control unnecessary and high-frequency feedback in the CTMBIDS, the model uses equations (\ref{eq4}) and (\ref{eq5}) for clause feedback activation function. In other words, when the activation probability decreases and the number of votes reaches the $T$ hyperparameter that had already been set, the feedback does not activate. This indicates that the TAs do not change their state anymore. This process makes room for remaining clauses to search for other sub-patterns. Therefore, the resource allocation can be done far more efficiently. This main principle also holds true for feedback type II. After all the models are trained, we start the model evaluation process. In the next section, we will provide the empirical results of the CTMBIDS model as well as other models that we implemented for comparison.

\begin{table}[h]
\caption{Hyperparameter values for the CTMBIDS model.}\label{tab9}
\begin{tabular}{ll} 
\toprule
Hyperparameter & Value \\
\midrule
Precision ($S$) & 3.9 \\
Threshold ($T$) & 20 \\
Number of Clauses ($m$) & 500 \\
Kernel size ($K$) & $3 \times 3$ \\
\bottomrule
\end{tabular}
\end{table}

As mentioned before, our proposed CTMBIDS method will be compared with 7 other models in order to have a benchmark of the proposed method. Table \ref{tab3} shows the details of these 7 models that we use to compare the CTMBIDS model with. In the following section, we will discuss the results of the proposed CTMBIDS method.

\section{Results Evaluation and Discussion}\label{sec5}
As we said before, this paper contains two main phases. One phase created 3 main SDN-based dataset that were discussed in detail. In the second phase, the preprocessing data yielded 16 features using the RFE method. Our experiment was implemented in two phases as well. Phase 1 of the experiment includes performance of the proposed CTMBIDS model along with 7 other state-of-the-art models. Phase 2 includes clause variation for the CTMBIDS, which shows the performance of the CTMBIDS model utilizing different clauses for training. The implementation details and codes are available on the github account [\href{https://github.com/russelljeffrey/CTMBIDS}{CTMBIDS github}]. 
\subsection{Phase 1: Experimental Configuration}
The models were implemented using Python programming language as well as a number of libraries that are extensively used in Python programming language. Table \ref{tab10} demonstrates the experimental configuration of our proposed method along with all the libraries that were used to design and implement all the models. 

\begin{table}[h]
\caption{Experimental configuration for CTMBIDS model.}\label{tab10}
\begin{tabular}{l|l} 
\toprule
Operating System & Windows 11 \\
Memory & 16.0 GB \\
CPU & Intel(R) Core(TM) i7-4720HQ CPU @ 2.60GHz, 2.60 GHz \\
Python & 3.9.13 \\
pyTsetlinMachine & 0.6.4 \\
Pandas & 2.1.1 \\
Numpy & 1.25.2 \\
Tensorflow & 2.14.0 \\
scikit-learn & 1.3.1 \\
memory\_profiler & 0.61.0 \\
\bottomrule
\end{tabular}
\end{table}

\subsection{Phase 1: Evaluation Metrics}
In order to evaluate all models, we used three of the most common evaluation metrics such as accuracy, precision and f1-score. Furthermore, we used memory\_profiler \cite{45} in order to compare the CTMBIDS model with other models in terms of its memory usage. The following equations demonstrate how accuracy, precision and f1-score are calculated:

\begin{equation}
Accuracy = \frac{TP + TN}{TP + TN + FP + FN}\label{eq7}
\end{equation}

\begin{equation}
Precision = \frac{TP}{TP + FP}\label{eq8}
\end{equation}

\begin{equation}
F1-Score = \frac{2 \times TP}{2 \times TP + FP + FN}\label{eq9}
\end{equation}
\\
In above equations, True Positive ($TP$) and True Negative ($TN$) represent the values that were correctly classified and False Positive ($FP$) and False Negative ($FN$) represent the values that were incorrectly predicted. We provide the analysis of our results in great detail. These results include the CTMBIDS model and 7 other models for comparative analysis. These 7 models will be used as a benchmark to compare the performance of CTMBIDS with other state of the art IDSs. These models were compared in detail in table \ref{tab3}. The algorithms that are used for each model are according to table \ref{tab3}, i.e, Logistic Regression \cite{38}, KNN \cite{34}, SVM \cite{35}, Random Forest \cite{33}, Naive Bayes \cite{36}, Tsetlin Machine \cite{40} and CNN-LSTM \cite{39}.

As we discussed in the feature selection section, our empirical results include both full features of the dataset as well as the 16 features that were chosen by the RFE algorithm. We will address the datasets with full features (Table \ref{tab6}) and selected features (Table \ref{tab8}) as full-feature datasets and sub-feature datasets, respectively. As we mentioned earlier in the Data Sampling section, we conducted our method using 10 fold cross validation. Each fold uses 10,000 samples of data, which indicates that 100,000 data were chosen from the dataset for training and testing. Table \ref{tab11} and table \ref{tab12} represent the accuracy, precision, f1-score and memory consumption of all models for sub-feature and full-feature datasets, respectively. 

The comparison for each model is carried out based on individual datasets. Thus, the bold value in each dataset column represents the highest value for accuracy, precision and f1-score for that specific dataset. Moreover, the lowest value for memory consumption in each dataset column is in bold to indicate the least memory consumption for each model in regards to a dataset. In table \ref{tab11}, it can be seen that Random Forest has the highest accuracy, precision and f1-score for Data18. In spite of Random Forest accuracy, its memory consumption is considerably high. It is clearly evident that the proposed CTMBIDS does not have the highest accuracy for sub-feature Data18, Data50 and Data100 datasets. However, when we compare the proposed model accuracy with the Random Forest model, the memory consumption is far lower. It seems unreasonable for the Random Forest model to consume 4 times the memory of the proposed CTMBIDS model in order to gain less than 1 percent more accuracy. For the KDDCup99 dataset however, the proposed CTMBIDS model shows to have the highest accuracy among other models in spite of using the least memory.

\begin{table}[]
\caption{Accuracy, Precision and F1-Score for sub-feature datasets.}\label{tab11}
\begin{tabular}{llllll}
\toprule
\multicolumn{1}{l}{Model Name}                & Evaluation Metrics (\%)  & Data18         & Data50         & Data100        & KDDCup99       \\
\midrule
\multirow{4}{*}{Logistic Regression \cite{38}} & Average Accuracy         & 97.99          & 97.97          & 97.26          & 97.51          \\
                                              & Average Precision        & 96.90          & 97.02          & 95.43          & 94.16          \\
                                              & Average F1-Score         & 98.10          & 98.49          & 97.42          & 93.76          \\
                                              & Memory Consumption (MiB) & 2526.7         & 2771.5         & 2778.7         & 2683.9         \\
\midrule
\multirow{4}{*}{KNN \cite{34}}                 & Average Accuracy         & 99.03          & 98.18          & 98.44          & 99.10          \\
                                              & Average Precision        & 98.35          & 97.33          & 97.24          & 99.21          \\
                                              & Average F1-Score         & 99.08          & 98.64          & 98.51          & 99.06          \\
                                              & Memory Consumption (MiB) & 1074.3         & 1196.8         & 1293.6         & 1208.0         \\
\midrule
\multirow{4}{*}{SVM \cite{35}}                 & Average Accuracy         & 98.62          & 98.11          & 98.44          & 95.53          \\
                                              & Average Precision        & 97.52          & 97.22          & 97.46          & 96.45          \\
                                              & Average F1-Score         & 98.70          & 98.59          & 98.53          & -              \\
                                              & Memory Consumption (MiB) & 1109.7         & 1127.3         & 1278.8         & 1199.9         \\
\midrule
\multirow{4}{*}{Random Forest \cite{33}}      & Average Accuracy         & \textbf{100}   & 99.95          & 99.99          & 99.88          \\
                                              & Average Precision        & \textbf{100}   & 99.91          & 99.99          & -              \\
                                              & Average F1-Score         & \textbf{100}   & 99.97          & \textbf{99.96} & -              \\
                                              & Memory Consumption (MiB) & 1365.5         & 1443.1         & 1484.5         & 1357.5         \\
\midrule
\multirow{4}{*}{Naive Bayes \cite{36}}         & Average Accuracy         & 97.50          & 95.46          & 96.90          & 98.50          \\
                                              & Average Precision        & 95.54          & 93.57          & 95.03          & -              \\
                                              & Average F1-Score         & 97.66          & 96.68          & 97.08          & -              \\
                                              & Memory Consumption (MiB) & 986.2          & 1149.9         & 1084.5         & 1006.8         \\
\midrule
\multirow{4}{*}{Tsetlin Machine \cite{40}}     & Average Accuracy         & 99.89          & 99.96          & 98.90          & 97.36          \\
                                              & Average Precision        & 99.93          & \textbf{99.97} & 99.31          & 98.28          \\
                                              & Average F1-Score         & 99.90          & 99.97          & 98.94          & 98.27          \\
                                              & Memory Consumption (MiB) & 1275.0         & 1128.3         & 1031.3         & 1262.0         \\
\midrule
\multirow{4}{*}{CNN-LSTM \cite{39}}            & Average Accuracy         & \textbf{100}   & \textbf{100}   & \textbf{100}   & 96.32          \\
                                              & Average Precision        & 99.98          & \textbf{99.97} & \textbf{99.99} & 95.39          \\
                                              & Average F1-Score         & 99.97          & \textbf{100}   & 99.95          & 95.51          \\
                                              & Memory Consumption (MiB) & 3879.9         & 3713.1         & 3812.7         & 3834.8         \\
\midrule
\multirow{4}{*}{Proposed CTMBIDS}             & Average Accuracy         & 98.28          & 99.89          & 98.16          & \textbf{99.98} \\
                                              & Average Precision        & 99.49          & 99.86          & 98.56          & \textbf{99.83} \\
                                              & Average F1-Score         & 98.23          & 99.92          & 98.20          & \textbf{99.80} \\
                                              & Memory Consumption (MiB) & \textbf{371.3} & \textbf{364.8} & \textbf{377.1} & \textbf{325.8}\\
\bottomrule
\end{tabular}
\end{table}

Table \ref{tab12}, illustrates the accuracy, precision and f1-score of all the models for full-feature datasets. For Data18, Logistic Regression, SVM and CNN-LSTM models have the highest accuracy, precision and f1-score compared to other models. Once again, Logistic Regression, SVM and CNN-LSTM show that they require more memory compared to the CTMBIDS model. In the remaining datasets, CTMBIDS excels far better than the remaining models. What is more interesting in case of full-feature datasets is that the proposed CTMBIDS can reasonably keep the memory consumption low while handling a high number of features in the dataset. This can be easily comprehended when we take the number of KDDCup99 features into account. The more features there are for the model to be trained with, the more memory the model consumes. This makes the CTMBIDS model more appropriate in a resource-constrained environment. This issue is particularly important in an SDN environment where there is no abundance of memory available for the anomaly-based IDS in an SDN controller.

\begin{table}[]
\caption{Accuracy, Precision and F1-Score for full-feature datasets.}\label{tab12}
\begin{tabular}{llllll}
\toprule
\multicolumn{1}{l}{Model Name}       & {Evaluation Metrics (\%)} & {Data18} & {Data50} & {Data100} & {KDDCup99} \\
\midrule
\multirow{4}{*}{Logistic Regression \cite{38}} & Average Accuracy                 & \textbf{100}    & 99.85           & 99.45            & 94.96             \\
                                              & Average Precision                & 99.97           & 99.71           & 99.61            & 87.98             \\
                                              & Average F1-Score                 & 99.93           & 98.83           & 99.73            & 87.32             \\
                                              & Memory Consumption (MiB)         & 10423.1         & 10923.5         & 10134.1          & 14322.8           \\
\midrule
\multirow{4}{*}{KNN \cite{34}}                 & Average Accuracy                 & 99.99           & 98.67           & 99.75            & 96.13             \\
                                              & Average Precision                & 99.93           & 98.90           & 99.56            & 97.54             \\
                                              & Average F1-Score                 & 99.98           & 99.00           & 99.76            & 95.72             \\
                                              & Memory Consumption (MiB)         & 2116.3          & 2307.5          & 2192.0           & 4261.0            \\
\midrule
\multirow{4}{*}{SVM \cite{35}}                 & Average Accuracy                 & \textbf{100}    & 99.22           & 99.80            & 92.16             \\
                                              & Average Precision                & \textbf{100}    & 99.51           & 99.67            & 93.77             \\
                                              & Average F1-Score                 & \textbf{100}    & 99.41           & 99.81            & 93.26             \\
                                              & Memory Consumption (MiB)         & 1879.3          & 1841.0          & 2068.2           & 3949.9            \\
\midrule
\multirow{4}{*}{Random Forest \cite{33}}       & Average Accuracy                 & 100             & 99.96           & 100              & 98.15             \\
                                              & Average Precision                & 99.84           & 99.91           & 99.96            & 97.36             \\
                                              & Average F1-Score                 & 99.95           & 99.97           & 99.95            & 97.02             \\
                                              & Memory Consumption (MiB)         & 2807.3          & 2457.1          & 2829.0           & 4779.8            \\
\midrule
\multirow{4}{*}{Naive Bayes \cite{36}}         & Average Accuracy                 & 99.99           & 95.45           & 98.12            & 97.66             \\
                                              & Average Precision                & \textbf{100}    & 99.91           & 96.51            & 96.81             \\
                                              & Average F1-Score                 & 99.99           & 96.43           & 98.22            & 95.53             \\
                                              & Memory Consumption (MiB)         & 3455.1          & 3403.1          & 3682.8           & 8671.4            \\
\midrule
\multirow{4}{*}{Tsetlin Machine \cite{40}}     & Average Accuracy                 & 99.99           & 99.45           & 99.73            & 99.45             \\
                                              & Average Precision                & 99.99           & 99.62           & 99.47            & 98.66             \\
                                              & Average F1-Score                 & 99.95           & 99.51           & 99.60            & 98.19             \\
                                              & Memory Consumption (MiB)         & 2569.1          & 2383.0          & 2478.8           & 3736.8            \\
\midrule
\multirow{4}{*}{CNN-LSTM \cite{39}}            & Average Accuracy                 & \textbf{100}    & 99.97           & 99.88            & 98.63             \\
                                              & Average Precision                & \textbf{100}    & 99.76           & 99.73            & 97.47             \\
                                              & Average F1-Score                 & \textbf{100}    & 99.82           & 99.79            & 94.84             \\
                                              & Memory Consumption (MiB)         & 6437.4          & 6504.3          & 6801.0           & 9651.5            \\
\midrule
\multirow{4}{*}{Proposed CTMBIDS}             & Average Accuracy                 & 99.97           & \textbf{99.99}  & \textbf{99.95}   & \textbf{99.43}    \\
                                              & Average Precision                & 99.96           & \textbf{99.98}  & \textbf{99.98}   & \textbf{99.65}    \\
                                              & Average F1-Score                 & 99.96           & \textbf{99.99}  & \textbf{99.97}   & \textbf{99.64}    \\
                                              & Memory Consumption (MiB)         & \textbf{1348.3} & \textbf{1523.1} & \textbf{1503.1}  & \textbf{1891.0}\\
\bottomrule
\end{tabular}
\end{table}

\begin{figure}[h]
\centering
\includegraphics[width=0.8\textwidth]{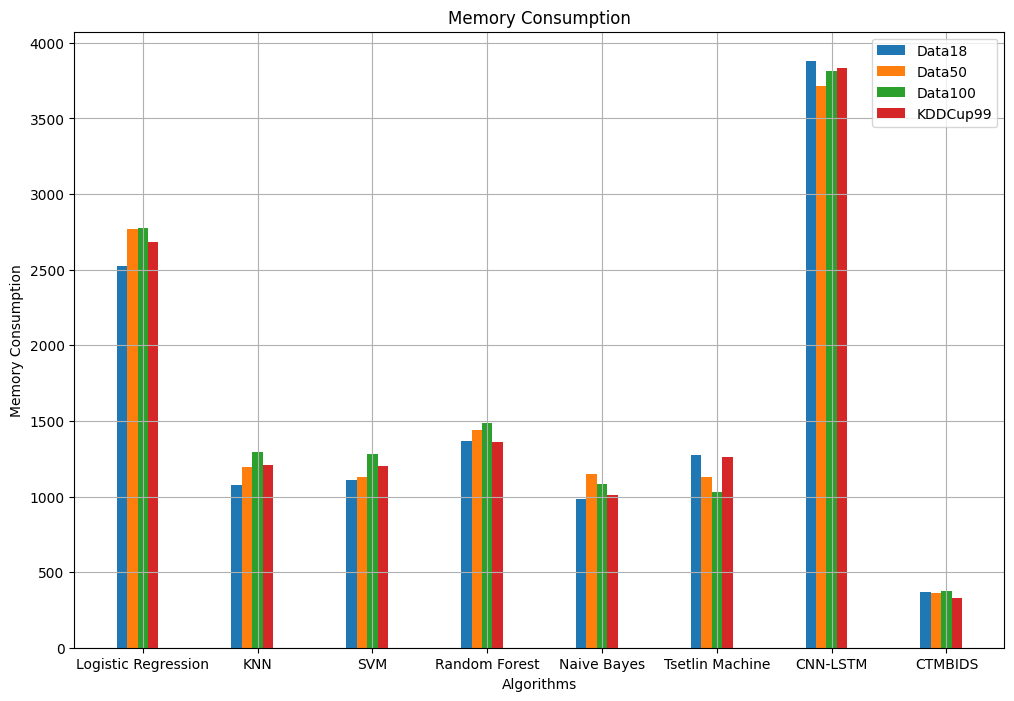}
\caption{Models' memory consumption for sub-feature datasets.}\label{fig9}
\end{figure}

\begin{figure}[h]
\centering
\includegraphics[width=0.8\textwidth]{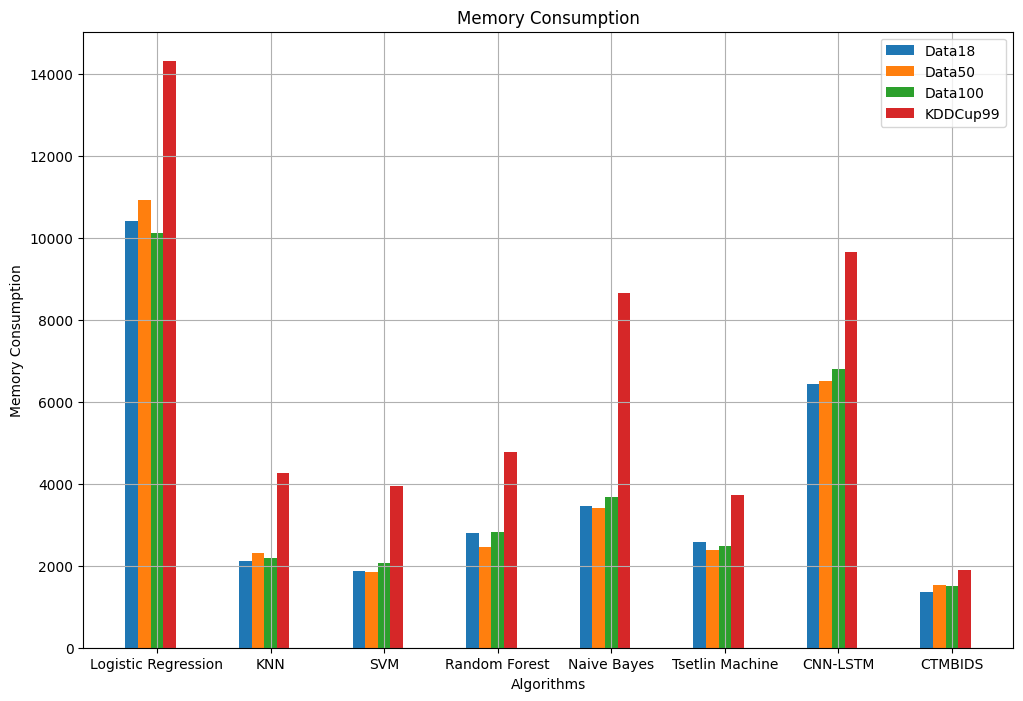}
\caption{Models' memory consumption for full-feature datasets.}\label{fig10}
\end{figure}

For a better understanding of models' memory consumption, we provided the memory consumption of all the models for both sub-feature and full-feature datasets in figure \ref{fig9} and \ref{fig10}, respectively. It is clear that the proposed CTMBIDS model consumes the least memory in all the datasets while CNN-LSTM and Logistic Regression have the highest memory consumption for sub-feature and full-feature datasets, respectively. This all indicates that the interpretability of the CTMBIDS model as well as its lightweight nature in memory consumption makes it a great approach for environments where resource abundance is not a reasonable option. These environments can be SDN environments, IoT and wireless networks. 

\subsection{Phase 2: Clause Variation}
\begin{figure}[h]
\centering
\includegraphics[width=0.8\textwidth]{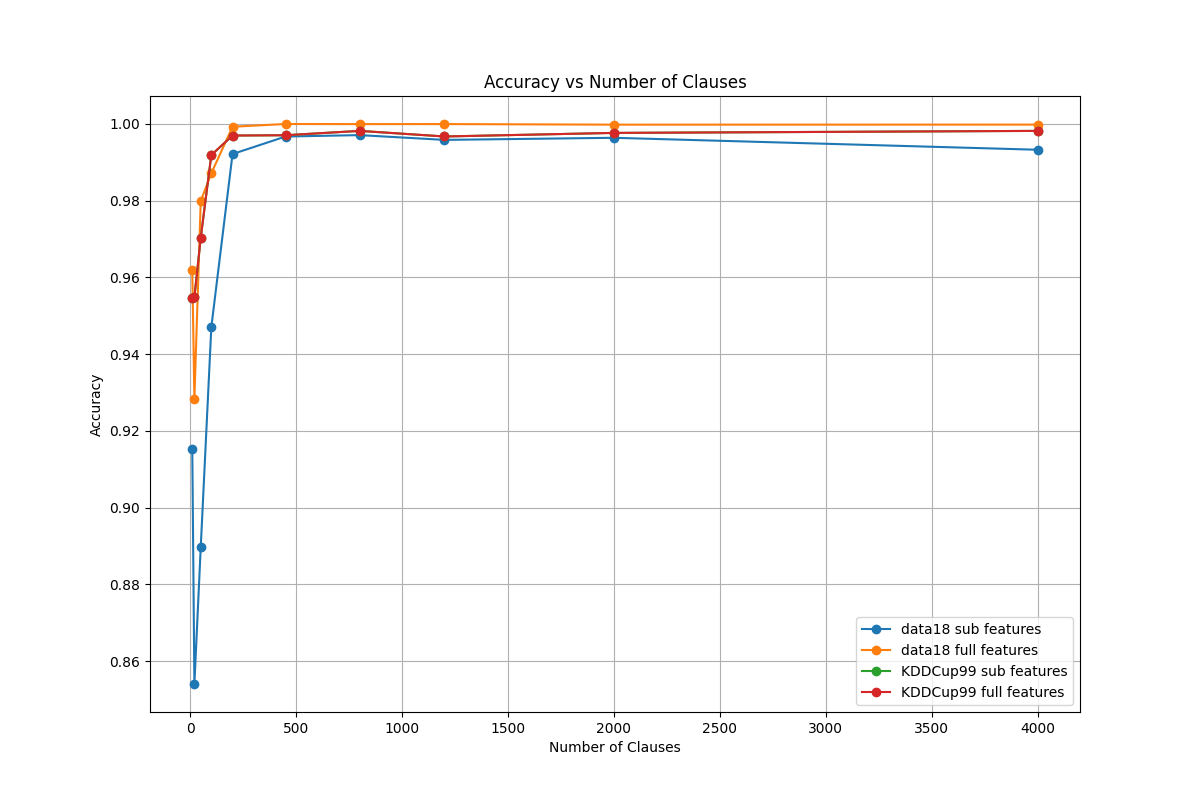}
\caption{CTMBIDS accuracy on Data18 and KDDCup99 using clause 10, 20, 50, 100, 200, 450, 800, 1200, 2000 and 4000.}\label{fig11}
\end{figure}

After implementation of all models, we implemented the CTMBIDS once more on Data18 and KDDCup datasets. This time we trained the model using a different number of clauses to see the model performance based on the number of clauses. Figure \ref{fig11} shows the accuracy of the CTMBIDS model on Data18 and KDDCup99 datasets both with sub-feature and full-feature datasets. As it can be seen, we used 10 different clauses for training the CTMBDS model. They are 10, 20, 50, 100, 200, 450, 800, 1200, 2000 and 4000. The steep curve in the figure shows that increasing the number of clauses from a certain point on does not increase the accuracy. In other words, the model overfits and the learning halts at a certain point. However, the dataset and the model architecture affect the model's learning curve and its accuracy as well. The more clauses are used in the model, the more memory the model consumes. Moreover, the more clauses are used in the model, the more time the model requires to complete the computations. Hence, it takes more time for the model to complete its training. Adjusting the "clause" hyperparameter is a necessity in order to maintain an ideal memory consumption and reasonable time complexity during the model training.

\section{Conclusion and Future Work}
The SDN environment is extremely beneficial for turning innovative ideas into applicable softwares. Its flexibility and programmability facilitate the growth of the network. However, this growth is coupled with threats and security risks that if not handled, can wreak havoc throughout the entire network and halt the widespread development of the SDN architecture. Thus, development and enhancement of IDSs using novel approaches can both fortify the SDN architecture from security flaws in the network and help network administrators more carefully handle the network management. In this work, we provided a lightweight IDS for detecting DDoS and DoS attacks in the SDN environments, which utilizes the least amount of memory compared to other state-of-the-art proposed approaches. This efficiency saves memory and makes sure the IDS can still maintain its high accuracy. Another important hurdle of the SDN environments was the lack of data compared to data in traditional networks. This work also addressed this issue by generating three new datasets that can be used to train various types of anomaly based IDSs in the SDN environment.  
This study has aimed to pave the way for lightweight models that can take advantage of Tsetlin and Convolutional Tsetlin Machine Algorithms. Not only it can be used for network security, but also for Image Processing tasks. In the future, we will focus on utilizing CTM algorithm on image processing tasks that can yield an acceptable accuracy compared to state-of-the-art models such as CNN. We also would like to enhance the performance of the CTM algorithm using evolutionary algorithms in order to obtain better accuracy. Finally, we would like to address the fact that wireless networks are in need to utilize low-power models where energy consumption is of top priority. Hence, our future work will also include models in wireless networks, in which we can take advantage of both TM  and CTM algorithms' lightweight classification properties. 

\section{Declarations}
\subsection{Competing Interests} The authors have no financial or proprietary interests in any material discussed in this article.
\subsection{Authors Contribution Statement} Author Rasoul Jafari Gohari collected the data in simulation in addition to conceptualizing and designing the IDS while all authors collaboratively structured the article, prepared the manuscript, analyzed data, reviewed and edited drafts, and finally approved the publication.
\subsection{Code availability} All the codes and implementations for each phase of this publication are available and easily accessible via Github (\href{https://github.com/russelljeffrey/CTMBIDS}{CTMBIDS Link}). 
\subsection{Ethical and Informed Consent} This research utilizes the KDDCup99 dataset, which is publicly available for academic use. We adhere to ethical principles throughout our research. We prioritize data privacy, responsible use, and avoiding potential harm, focusing our methods on modeling based on KDDCup99 dataset without any unethical irresponsibility.

\end{document}